\documentclass[a4paper,11pt]{article}

\usepackage{amsmath}

\newcommand{\egc}{\mbox{e.\,g.\,}}

\newcommand{\vctr}[1]{\ensuremath{\mathbf{ #1 }}}

\newcommand{\pb}[2]{\ensuremath{\left\{ #1 , #2 \right\} }}

\newcommand{\dr}[1]{\ensuremath{\mathrm{d} #1\,}}
\newcommand{\mc}[1]{\ensuremath{\mathcal{#1}}}

\newcommand{\ddt}{\ensuremath{\frac{\dr{}}{\dr{t}}}}

\newcommand{\pbp}[2]{\ensuremath{\frac{\partial #1}{\partial #2}}}

\newcommand{\vbv}[2]{\ensuremath{\frac{\delta #1}{\delta #2}}}

\newcommand{\mtr}[4]{\ensuremath{\left( \begin{array}{cc} #1 & #2 \\
#3 & #4 \end{array} \right) }}

\newcommand{\be}{\begin{equation}}
\newcommand{\ee}{\end{equation}}

\newcommand{\re}{\mathrm{R}}
\renewcommand{\vctr}[1]{\ensuremath{\vec{#1}}}
\begin{document}

\title{First-class constraints generate gauge transformations in electromagnetism (reply to Pitts)}
\author{Oliver Pooley and David Wallace}
\maketitle

\begin{abstract}
Brian Pitts has recently claimed to show via straightforward calculation that, at least in the case of Hamiltonian electromagnetism, an arbitrary first-class constraint ``generates not a gauge transformation, but a bad physical change'' (\emph{Annals of Physics} 351 (2014) pp.382--406). %We demonstrate that this is not the case, in two senses. Two relevant notions of gauge transformation need to be distinguished: transformations of \emph{states}, and transformations of \emph{histories}. 
We show, via a straightforward calculation, that a transformation generated by an arbitrary first-class constraint relates gauge-equivalent phase space points, vindicating orthodoxy. Pitts, however, is primarily concerned with transformations of entire histories, rather than of instantaneous states. We show that, even in this context, a transformation generated by an arbitrary first-class constraint is also a gauge transformation, once the empirically observed electric field is correctly identified via its dynamical interactions with charge, and not simply given stipulatively as a certain combination of the potential and its derivatives.
\end{abstract}

\section{Introduction}

According to conventional wisdom, in the context of the constrained Hamiltonian formalism, an arbitrary first-class constraint generates gauge transformations, at least for well known, physically reasonable theories such as electromagnetism.\footnote{Recall that the first-class constraints are those constraints that have vanishing Poisson brackets with all the constraints \cite{diracconstraint}. The restriction to ``physically reasonable'' theories is intended to exclude well-known counterexamples of the type considered in, e.g., \cite{henneauxteitelboim}, pp~19--20. Our defence of the letter of orthodoxy for theories such as electromagnetism should not be taken to indicate that we also endorse such a position for totally constrained theories, such as general relativity. The issues here are subtle, for reasons explored in, e.g., \cite{wallaceBC,pooleyHR,barbourfoster,thebault3}.} In a provocative recent paper~\cite{pittsEM}, Brian Pitts has challenged this conventional wisdom:
\begin{quote}
For Maxwell's electromagnetism, where everyone knows what a gauge transformation is -- what makes no physical difference, namely, leaving \vctr{E} and \vctr{B} unchanged -- and where all the calculations are easy, one can \emph{test} the claim that a first-class constraint generates a gauge transformation. \ldots Surprisingly, given the age of the claim, such a test apparently has not been made before, at least not completely and successfully, and has rarely been attempted. Perhaps the temptation to default to prior knowledge has been irresistible.\ldots Anyway the test can be made \ldots The result is clearly negative: a first-class constraint -- either the primary or the secondary -- generates a physical difference, a change in \vctr{E}. This change involves the gradient of an arbitrary function, implying that $\nabla \cdot \vctr{E}\neq 0$, spoiling Gauss's law. 
\end{quote}

It would indeed be surprising (to say the least) if the conventional view %, that electromagnetic constraints generate gauge transformations,%
 could be shown to be wrong via an easy calculation that somehow had never been done explicitly. In this paper we shall argue that Pitts is mistaken. Pitts's calculations do not undermine the claim that the electromagnetic constraints generate gauge transformations.
 
Our argument has two stages, related to two natural senses in which the constraints might be said to generate gauge transformations: (i) the transformations might always map a phase space point to a point which represents the same instantaneous state and (ii) they might always map a dynamical curve through phase space to a curve that represents the same physical history. % So we omit all mention of the other two notions you discuss in the original draft?
 Pitts's argument targets only the claim that first-class constraints generate gauge transformations in the second of these senses, and yet he presents his argument as a refutation of the conventional wisdom. This is odd because the authors he cites as exemplars of conventional wisdom (e.g., \cite{diracconstraint,henneauxteitelboim}) are in the first instance concerned with gauge transformations in the first sense. We show that, in electromagnetism, phase space points connected by transformations generated by arbitrary first-class constraints always do correspond to the same physical state, just as conventional wisdom requires. (In fact this is a corollary of Pitts's own calculations, although one he does not draw.)

The second stage of our argument engages directly with Pitts's claim that only a proper subset of the first class constraints generate gauge transformations between histories. Pitts is correct that the full set of transformations generated by arbitrary first-class constraints is not even a symmetry group (and so \emph{a fortiori} not a gauge symmetry group) of `standard' Lagrangian-equivalent Hamiltonian electromagnetism. All such transformations are, however, symmetries of the so-called \emph{extended Hamiltonian formalism}. Pitts claims that this formalism is not physically equivalent to the Lagrangian theory, and that transformations generated by arbitrary first-class constraints relate phase space curves that, in general, correspond to empirically distinct histories. The key question in this context is: which dynamical object in the extended Hamiltonian formalism plays the role of the electric field; that is, which object couples to charge in the usual way. We demonstrate, \emph{contra} Pitts's claims, that it is the momentum conjugate to \vctr{A} that plays this role, and which therefore represents the familiar electric field. On this understanding, solutions connected by transformations generated by arbitrary combinations of the primary and secondary constraints are gauge-related.

The scope of Pitts's rich paper is wide-ranging, both conceptually and historically; this reply, by contrast, is intentionally narrow. It does not engage with, nor seek to challenge, the bulk of Pitts's observations, but is focussed specifically on Pitts's claim that the electromagnetic constraints do not generate gauge transformations, and that they can be shown not to do so by straightforward calculation. We claim no originality for any of the calculations in the present paper; for references to the formalism used in what follows, see, \egc, \cite{diracconstraint}, \cite{henneauxteitelboim} or \cite{matschull}.

\section{The Constrained-Hamiltonian Form of Electromagnetism}

The Lagrangian of electromagnetism, with respect to an arbitrarily chosen set of standard-simultaneity coordinates, is
\be  \label{lag}
L[\vctr{A},V;\dot{\vctr{A}},\dot{V}]=\int \left\{\frac{1}{2}(\dot{\vctr{A}}-\nabla V)^2 -\frac{1}{2}(\nabla \times \vctr{A})^2 -( V \rho + \vctr{A} \cdot \vctr{J})\right\}.
\ee
Here:
\begin{itemize}
\item The %coordinates
configuration variables are the vector field \vctr{A} and the scalar field $V$ (both taken to be time-dependent functions on $\re^3$ satisfying appropriate differentiability and boundary conditions);
\item Given a vector $\vctr{V}$, $V^2$ is shorthand for $\vctr{V}\cdot \vctr{V}$;
\item The integral is over $\re^3$ and coordinates are suppressed for readability;
\item $\vctr{J}$ and $\rho$  jointly describe a background charge distribution, introduced for later convenience and satisfying $\dot{\rho}+\nabla \cdot \vctr{J}=0$.
\end{itemize}
The Euler-Lagrange equations for this Lagrangian are:
\begin{eqnarray}
\ddt{(\dot{\vctr{A}}-\nabla V)} - \nabla \times (\nabla \times \vctr{A}) + \vctr{J}& =& 0 \label{lagdyn}\\
\nabla \cdot (\dot{\vctr{A}}-\nabla V)-\rho& =& 0. \label{lagcon}
\end{eqnarray}
For further convenience, we introduce $\vctr{B}$ as a definitional abbreviation for $\nabla \times \vctr{A}$. (We do \emph{not} at this point introduce \vctr{E} as a similar abbreviation for $\dot{\vctr{A}}-\nabla V$, for reasons that will become apparent.) %%added "at this point" because we do later - in equations 33 and 42, and in their preamble, immediately following the long Pitts quote in the extended Hamiltonian section

These equations have a time-dependent symmetry --  the usual gauge symmetry --  given by
\be 
V(\vctr{x},t)\rightarrow V(\vctr{x},t)+\pbp{\Lambda}{t}(\vctr{x},t);\,\,\,\,\,\vctr{A}(\vctr{x},t)\rightarrow \vctr{A}(\vctr{x},t)+\nabla \Lambda(\vctr{x},t)
\ee
for arbitrary smooth $\Lambda$.

If we calculate the Legendre transforms of $\dot{\vctr{A}}$ and $\dot{V}$, we find that
\be \label{legendre}
\vbv{L}{\dot{\vctr{A}}}= \dot{\vctr{A}}-\nabla V;\,\,\,\,\, \vbv{L}{\dot{V}}=0.
\ee
This is clearly non-invertible and so we expect a constrained Hamiltonian. Applying the usual machinery of the constrained-Hamiltonian formalism yields the Hamiltonian
\be
H[\vctr{A},V;\vctr{\pi},\pi_0]=\int \frac{1}{2}\left\{(\vctr{\pi}^2 + \vctr{B}^2) + \lambda \pi_0 + \vctr{\pi}\cdot \nabla V + (V \rho + \vctr{A}\cdot \vctr{J})\right\}
\ee
and an infinite family of primary constraints
\be 
C_0(\vctr{x}) \equiv \vctr{\pi}_0(\vctr{x}) =0
\ee
where:
\begin{itemize}
\item \vctr{\pi} and $\pi_0$ are the momenta conjugate to $\vctr{A}$ and $V$ respectively;
\item $\lambda$ is an arbitrarily chosen Lagrange multiplier;
\item the Poisson brackets take the usual form (restoring vector component indices and coordinates)
\be 
\pb{A^i(\vctr{x})}{\pi_j(\vctr{y})}= \delta^i_j \delta(\vctr{x}-\vctr{y});\,\,\,\,\pb{V(\vctr{x})}{\pi_0(y)}=\delta(\vctr{x}-\vctr{y});\,\,\,\, \text{all others vanish}
\ee
where $\delta^i_j$ is the Kronecker delta and $\delta(\vctr{x}-\vctr{y})$ is the Dirac delta function.
\end{itemize}
Integrating by parts (assuming appropriate boundary conditions, so that the fields vanish sufficiently rapidly at infinity) allows us to rewrite the Hamiltonian as
\be 
H[\vctr{A},V;\vctr{\pi},\pi_0]=\int \frac{1}{2}(\vctr{\pi}^2 + \vctr{B}^2) +  \vctr{A}\cdot \vctr{J} - V(\nabla \cdot \vctr{\pi} -  \rho) + \lambda \pi_0.
\ee
We find
\be 
\pb{H}{C_0(\vctr{x})} = - \vbv{H}{V(\vctr{x})}= - (\nabla \cdot \vctr{\pi}(\vctr{x}) -  \rho(\vctr{x})).
\ee
So there is also an infinite family of \emph{secondary} constraints
\be 
C_1(\vctr{x})=\nabla \cdot \vctr{\pi}(\vctr{x})- \rho(\vctr{x}).
\ee
As usual in constrained Hamiltonian mechanics, the primary constraint arises because the Legendre transform is not 1-to-1: phase-space points not on the primary constraint surface do not correspond to any Lagrangian state. The secondary constraint arises from (indeed, is the Legendre transform of) the Lagrangian-formalism constraint equations (\ref{lagcon}): phase-space points not on the secondary constraint surface (but on the primary constraint surface) correspond to Lagrangian states not permitted by the Euler-Lagrange equations.

Written in terms of the constraints, the Hamiltonian becomes
\begin{eqnarray} \label{hamilt}
H[\vctr{A},V;\vctr{\pi},\pi_0]&=& \int \left\{\frac{1}{2}(\vctr{\pi}^2 + \vctr{B}^2) +  \vctr{A}\cdot \vctr{J}+ \lambda C_0 - V C_1 \right\}\nonumber \\ &\equiv& H_0 [\vctr{A},\vctr{\pi}]+ \int \left(\lambda C_0 - V C_1\right).
\end{eqnarray}
The equations of motion can be calculated using the Poisson brackets:
\begin{eqnarray}\label{hamilteqn}
\pbp{\vctr{A}(\vctr{x})}{t}=\vbv{H}{\vctr{\pi}(\vctr{x})}& = &\vctr{\pi}(\vctr{x}) + \nabla V(\vctr{x}) \\
\pbp{\vctr{\pi}(\vctr{x})}{t}=-\vbv{H}{\vctr{A}(\vctr{x})}& = &- \nabla \times \vctr{B}(\vctr{x}) - \vctr{J}(\vctr{x}) \label{hamilteqn2}\\
\pbp{V(\vctr{x})}{t}=\vbv{H}{\pi_0(\vctr{x})}&=& \lambda(\vctr{x})\label{hamilteqn3}\\
\pbp{\pi_0(\vctr{x})}{t}=-\vbv{H}{V(\vctr{x})}& = & C_1(\vctr{x}) = 0.\label{hamilteqn4}
\end{eqnarray}
As a useful corollary, we have
\be \label{Beqn}
\pbp{\vctr{B}(\vctr{x})}{t}=\nabla \times \vctr{\pi}(\vctr{x}).
\ee

\section{Gauge transformations of the state}

In Lagrangian electromagnetism, the instantaneous configuration of the system is given by a pair $(\vctr{A},V)$; if a given pair is the state of the sytem at time $t$, it transforms under a gauge transformation specified by function $\Lambda(\vctr{x},t)$ as follows:
\be \vctr{A}(\vctr{x})\rightarrow \vctr{A}(\vctr{x})+\nabla \Lambda(\vctr{x},t);\,\,\,\, V(\vctr{x})\rightarrow V(\vctr{x})+\dot{\Lambda}(\vctr{x},t).
\ee
If we calculate the induced change of the momenta, via the Legendre transformation %"tranform" (singular) didnt seem right
 (\ref{legendre}), we get
\be 
\vctr{\pi}(\vctr{x})\rightarrow \vctr{\pi}(\vctr{x});\,\,\,\,\pi_0(\vctr{x})\rightarrow \pi_0(\vctr{x})=0.
\ee
At any given time $t$, $\Lambda(\vctr{x},t)$ and $\dot{\Lambda}(\vctr{x},t)$ are independent of one another, so we have a criterion for gauge equivalence of Hamiltonian states:
\be \label{stategauge}
(\vctr{A},V,\vctr{\pi},\pi_0) \mbox{ is gauge-equivalent to }(\vctr{A}+\nabla g,V+f,\vctr{\pi},\pi_0)\mbox{ for arbitrary }f,g.
\ee
But now consider the phase-space function
\be 
O[f,g]=\int (f C_0 - g C_1)= \int (f \pi_0 +\vctr{\pi}\cdot \nabla g).
\ee
We can easily calculate that
\be
\pb{\vctr{A}}{O[f,g]}=\nabla g;\,\,\,\,\pb{V}{O[f,g]}=f;\,\,\,\,\pb{\vctr{\pi}}{O[f,g]}=\pb{\pi_0}{O[f,g]}=0.
\ee
So the constraints generate exactly the transformations of form (\ref{stategauge}). Thus the constraints generate gauge transformations in the first sense of gauge transformation described above: \emph{arbitrary linear combinations of the primary and secondary constraints generate transformations between gauge-related instantaneous states}.

\section{The extended Hamiltonian}

As noted, however, Pitts does not have this sense of gauge transformations in mind. His position is that arbitrary combinations of the primary and secondary constraints (and, in particular, the primary or secondary constraints by themselves) do not generate gauge transformations \emph{when acting on histories}. Translating his argument into our notation: let $f(\vctr{x},t)$ and $g(\vctr{x},t)$ be arbitrary functions of space and time, and write $f_t(\vctr{x})\equiv f(\vctr{x},t)$, $g_t(\vctr{x})\equiv g(\vctr{x},t)$. Then the infinitesimal action of the time-dependent function on phase space
\be O_t[f,g] \equiv O[f_t,g_t]\equiv \int f_tC_0-g_tC_1\ee
on the configuration variables $\vctr{A}$, $V$ is
\[
\vctr{A}(\vctr{x},t)\rightarrow\vctr{A'}(\vctr{x},t)= \vctr{A}(\vctr{x},t)+\epsilon\left\{\vctr{A}(\vctr{x},t),O[f_t,g_t]\right\}=\vctr{A}(\vctr{x},t)+\epsilon\nabla g(\vctr{x},t)
\]
\be
V(\vctr{x},t)\rightarrow V'(\vctr{x},t)= V(\vctr{x},t) + \epsilon\left\{V(\vctr{x},t),O[f_t,g_t]\right\}=V(\vctr{x},t) + \epsilon f(\vctr{x},t).
\ee

Now suppose $(\vctr{A},V)$, prior to the transformation, satisfy the Euler-Lagrange equations (\ref{lagdyn}--\ref{lagcon}). The transformed quantities $(\vctr{A}',V')$ then satisfy
\begin{eqnarray}
\ddt{(\dot{\vctr{A'}}-\nabla V')} - \nabla \times \vctr{B}' + \vctr{J}& =& \epsilon \ddt{\nabla(\dot{g}-f)} \\
\nabla \cdot (\dot{\vctr{A'}}-\nabla V')-\rho&=&\epsilon \nabla^2(\dot{g}-f).
\end{eqnarray}
So only in the special case where $\dot{g}-f$ is independent of $x$ is the form of the Euler-Lagrange equations preserved; in other cases, the transformation is not even a symmetry (and so not the standard gauge symmetry). In particular, the transformations generated by $C_0$ alone and by $C_1$ alone never preserve the form of the equations. Since Hamilton's equations reproduce the Euler-Lagrange equations, it also follows that the constraint-transformed histories fail to solve Hamilton's equations, and hence are not symmetries of the Hamiltonian formalism.\footnote{This, indeed, is the uncontroversial moral of Pitts's central calculations. We note that it is a fact remarked upon, not just in the literature focused on Hamiltonian--Lagrangian equivalence, and favourably cited by Pitts, but also by the authors he criticises: e.g., \cite{henneauxteitelboim}, Ch.~3.}

There is --- as is well known, and as Pitts discusses --- a modification of the Hamiltonian which does have these transformations as symmetries. The so-called ``extended Hamiltonian'' is obtained from the original Hamiltonian by adding an arbitrary multiple of the \emph{secondary} constraints:
\be 
H_{ext}[\vctr{A},V;\vctr{\pi},\pi_0]=H_0[\vctr{A},\vctr{\pi}]+ \int \left(\lambda C_0 - (V+\mu') C_1\right).
\ee
Here $\mu'$ is an arbitrary Lagrange multiplier. The term $V+\mu'$ is thus also arbitrary, and we can combine them into a single term, $\mu=V+\mu'$.

It will be useful to write the equations of motion generated by $H_{ext}$ explicitly:
\begin{eqnarray}\label{hamiltext}
\pbp{\vctr{A}(\vctr{x})}{t}=\vbv{H_{ext}}{\vctr{\pi}(\vctr{x})}& = &\vctr{\pi}(\vctr{x}) + \nabla \mu(\vctr{x}) \\
\pbp{\vctr{\pi}(\vctr{x})}{t}=-\vbv{H_{ext}}{\vctr{A}(\vctr{x})}& = &- \nabla \times \vctr{B}(\vctr{x}) - \vctr{J}(\vctr{x}) \label{hamiltext2}\\
\pbp{V(\vctr{x})}{t}=\vbv{H_{ext}}{\pi_0(\vctr{x})}&=& \lambda(\vctr{x})\label{hamiltext3}\\
\pbp{\pi_0(\vctr{x})}{t}=-\vbv{H_{ext}}{V(\vctr{x})}& = &  0.\label{hamiltext4}
\end{eqnarray}
We can swiftly verify that the transformations generated by $O_t[f,g]$ are symmetries of \emph{these} equations for arbitrary $f$, $g$. Furthermore, taking the curl of (\ref{hamiltext}) yields
\be 
\pbp{\vctr{B}(\vctr{x})}{t}=\nabla \times \vctr{\pi}(\vctr{x})
\ee
which, together with (\ref{hamiltext2}), the secondary constraints $\nabla \cdot \vctr{\pi}=\rho$, and the identity $\nabla \cdot \vctr{B}=0$, forms a self-contained set of equations for $\vctr{\pi}$ and $\vctr{B}$ which are recognisable as the Maxwell equations, with $\vctr{\pi}$ playing the role of the electric field.

There is an obvious temptation: conclude that $\vctr{\pi}$ \emph{is} the electric field, and deduce that while the extended-Hamiltonian theory is not literally Lagrangian electrodynamics, it is (at least) %added the parenthesis because the equivalence surely goes beyond the merely empirical
empirically equivalent to it. Pitts anticipates and rejects this temptation:

\begin{quote}
While it is acknowledged that the extended Hamiltonian is not equivalent to $L$ strictly, this inequivalence is often held to be harmless because they are equivalent for “observables.” %This claim presumably is intended to mean that the extended Hamiltonian is empirically equivalent to $L$, differing only about unobservable matters. Such a response will be satisfactory only if ``observable'' here is used in the ordinary sense of running experiments. Technical stipulations about the word ``observable,'' especially distinctively Hamiltonian stipulations, are irrelevant. 
\ldots Unfortunately it is not the case that the extended Hamiltonian is empirically equivalent to the Lagrangian, a fact that has been masked by equivocating on the word “observable” between the ordinary experimental sense and a technical Hamiltonian sense. It is peculiar to think of observing canonical momenta conjugate to standard Lagrangian coordinates --- in fact it seems to be impossible to observe that kind of canonical momentum as such. What would be the operational procedure for observing $p_i$? Rather, its experimental significance is purely on-shell, parasitic upon the observability of suitable functions of $q^i$ and/or derivatives of $q^i$ --- derivatives (spatial and temporal) of [$\vctr{A}$ and $V$] in the electromagnetic case. 
\end{quote}
(here and below we modify Pitts' notation to match ours.)

Pitts' position here is that the empirically accessible electric field should by understood to be represented by $\dot{\vctr{A}}-\nabla V$ and not by $\vctr{\pi}$ when the two differ; hence, what should be required for empirical equivalence is preservation of the equations of motion for $\vctr{E} \equiv \dot{\vctr{A}}-\nabla V$ and $\vctr{B}$, not $\vctr{\pi}$ and $\vctr{B}$. He gives two sets of reasons for this. The first is conceptual: (i) the canonical momentum ``does not even  appear as an independent field in the Lagrangian formalism, which formalism is correct and transparent'' and (ii) ``canonical momenta are auxiliary fields in the Hamiltonian action''. But (i) seems question-begging, since the issue is precisely whether a given modification of the Hamiltonian formalism is acceptable even though it fails to reproduce the Lagrangian form of the theory. And (ii) could apply equally to $V$, which is an auxiliary field in the Lagrangian formalism, eliminable in principle by solving the constraint. 

The second set of reasons is more interesting and gets at the heart of what it is for something to be ``observable'' in the empirical sense:

\begin{quote}
[T]he electric field is what pushes on charge; but it is easy to see that in both the Lagrangian and Hamiltonian contexts, what couples to the current density is not [$\vctr{\pi}$], but [$\vctr{A}$ and $V$]. \ldots What is the operational procedure for measuring $p_i$? The only plausible answer is to use on-shell equivalence to the empirically available [$\vctr{E}$], which involves derivatives of $[\vctr{A}$ and $V$]. Otherwise, what reason is there to believe that any procedure for measuring $[\vctr{\pi}]$ involves a measurement of the quantity that pushes on charge? 
\end{quote}
Pitts is surely correct that the operational significance of the electric field is entirely tied up with its dynamical relation to charge: with the way in which electric fields are generated by, and induce accelerations in, charges and currents. But he is incorrect that this argument speaks in favour of regarding $\dot{\vctr{A}}-\nabla V$ rather than $\vctr{\pi}$ as the electric field, if the dynamics in question are those generated by the extended Hamiltonian.

We can see this already in the background-charge version of electromagnetism which we have been considering. The Maxwell equations determine how electromagnetic fields are generated by charges, and we have seen that the extended-Hamiltonian formalism yields the Maxwell equations for $\vctr{\pi}$ and $\vctr{B}$. It is the field lines of $\vctr{\pi}$, not $\vctr{E}$, that have positive and negative charges as their sources and sinks.

What of the action of the fields on charge? For a quick-and-dirty route to these (for which apologies to readers allergic to the delta function), consider a charged particle with position $\vctr{X}(t)$ and couple it to the electromagnetic field via the Lagrangian
\be 
L[\vctr{A},V,X;\dot{\vctr{A}},\dot{V},\dot{X}]=\int \frac{1}{2}(\vctr{E}^2-\vctr{B}^2) + \frac{1}{2}m\dot{\vctr{X}}^2 - q( \vctr{A}(\vctr{x})\cdot\dot{\vctr{X}}  + V(\vctr{x})).
\ee
Calculating the Legendre transform, we find (since no time derivatives of $\vctr{A}$ or $V$ appear in the interaction term) that the electromagnetic momenta are as before, and that the vector $\vctr{P}$ of momenta conjugate to $\vctr{X}$ is given by
\be 
\vctr{P}=m \dot{X}-q \vctr{A}(\vctr{X}).
\ee
The Hamiltonian is then
\begin{eqnarray}
H[\vctr{A},V,\vctr{X};\vctr{\pi},\pi_0,\vctr{P}]&=&\int \left\{\frac{1}{2}(\vctr{\pi}^2 + \vctr{B}^2) + \lambda \pi_0  + \vctr{\pi}\nabla \cdot V\right\}\nonumber \\ &+&\frac{1}{2m}(\vctr{P}+q\vctr{A}(\vctr{X}))^2 + q V(\vctr{X})
\end{eqnarray}
with primary and secondary constraints 
\be 
0=C_0(\vctr{x})\equiv \pi_0(\vctr{x})\;\,\,\,\, 0=C_1(\vctr{x})\equiv \nabla \cdot \vctr{\pi}(\vctr{x})-q \delta(\vctr{x}-\vctr{X}).
\ee

As before we can pass to the extended Hamiltonian formalism by replacing $V$ with a Lagrange multiplier $\mu$; the equations of motion for $\vctr{A}$, $V$, $\vctr{\pi}$ and $\pi_0$ are exactly as in the background case and the equations of motion for the particle are
\begin{eqnarray} 
\dot{X}^i = \pbp{H_{ext}}{P_i} &=& \frac{1}{m}(P_i + q A_i(\vctr{X})) \label{particleX}\\
\dot{P}_i = -\pbp{H_{ext}}{X^i} &=& -\frac{q}{m}(\vctr{P}+q \vctr{A}(\vctr{X}))\cdot \pbp{\vctr{A}}{X^i} - q \pbp{\mu(\vctr{X})}{X^i}\nonumber \\
& =& -q\dot{\vctr{X}}\cdot \pbp{\vctr{A}}{X^i} - q \pbp{\mu(\vctr{X})}{X^i} \label{particleP}
\end{eqnarray}
Differentiating (\ref{particleX}), and substituting in (\ref{particleP} and \ref{hamiltext}) gives
\be 
m \ddot{\vctr{X}} =q \dot{\vctr{X}}\times \vctr{B}(\vctr{X}) + q (\dot{\vctr{A}}(\vctr{X})- \nabla \mu(\vctr{X})) = q (\dot{\vctr{X}}\times \vctr{B} + \vctr{\pi}),
\ee
which may be recognised as the Lorentz force law with $\vctr{\pi}$ playing the role of the electric field. So \emph{pace} Pitts, in this case at least it is the momentum $\vctr{\pi}$ which is ``the quantity that pushes on charge'' (and also has charge as its source and sink) and hence which is measured empirically when we determine the electric field.

For a more sophisticated case, we can consider coupling the electromagnetic field to a complex charged matter field $\phi$, via the Lagrangian
\be
L[\vctr{A},V,\phi;\dot{\vctr{A}},\dot{V},\dot{\phi}]=\int \left(\frac{1}{2}(\vctr{E}^2-\vctr{B}^2) +\frac{1}{2}(|\dot{\phi}- i V \phi |^2 - |\vctr{D}\phi|^2 -m^2|\phi|^2)\right)
\ee
where $\vctr{D}\equiv \nabla - i \vctr{A}$. To avoid some of the delicate issues that can arise with complex coordinates, we will simply treat $\phi$ as a two-component real field, replace multiplication by $i$ by action of the 2x2 matrix
\be 
\mc{I}=\mtr{0}{1}{-1}{0}
\ee
and treat $|\phi|^2$ as shorthand for the real inner product $\phi \cdot \phi$. The matter-field momentum $\psi$ is given by the Legendre transform:
\be 
\psi= \pbp{L}{\dot{\phi}}= \dot{\phi}- V \mc{I} \phi.
\ee
If we then define the charge and current as
\be 
\rho=\psi \cdot \mc{I}\phi; \,\,\,\, \vctr{J}= \nabla \phi \cdot \mc{I} \phi
\ee
(note that in the usual complex-field notation, $\alpha \cdot \mc{I} \beta = i(\alpha \beta^*-\beta \alpha^*)$)
we obtain as Hamiltonian
\begin{eqnarray}
H[\vctr{A},V,\phi;\vctr{\pi},\pi_0,\psi]&=&\int\frac{1}{2}(\vctr{\pi}^2 + \vctr{B}^2)\nonumber \\
& +& \int \frac{1}{2}(|\psi|^2 +|\vctr{D}\phi|^2 + m^2 |\phi|^2) \nonumber \\
& + & \int (- V(\nabla\cdot \vctr{\pi}-\rho) + \lambda \pi_0) 
\end{eqnarray}
with the usual primary and secondary constraints. We can then move to the extended Hamiltonian by replacing $V$ with the Lagrange multiplier $\mu$. The equations of motion are the familiar electromagnetic equations together with
\begin{eqnarray} 
\dot{\psi}=- \vbv{H_{ext}}{\phi}&=& - (-\vctr{D}^2+m^2)\phi +\mu \mc{I}\psi \\
\dot{\phi}=\vbv{H_{ext}}{ \psi} &=& \psi+ \mu \mc{I} \phi
\end{eqnarray}
from which we derive
\be 
\left(\ddt{} -\mc{I} \mu\right)^2 \phi - \vctr{D}^2\phi+m^2 \phi=0.
\ee
This is the standard equation of motion for a complex field, but with respect to the four-potential $(\vctr{A},\mu)$, not $(\vctr{A},V)$. When we recall also that $\vctr{\pi}$ satisfies $\vctr{\pi}=\dot{\vctr{A}}-\nabla \mu$, we see that it is the Lagrange multiplier $\mu$, not the coordinate $V$, that is playing the role of the scalar potential.

(Pitts actually considers the charged-scalar-field case (section 9), and writes ``The absence of terms connecting $[\phi$] with derivatives of $[\vctr{A},V]$ implies that charge couples to $[\vctr{A},V]$ and/or [their] derivatives, not to the canonical momenta conjugate to [$\vctr{A}$ and $V$], \emph{even in the Hamiltonian context}'' (emphasis ours). It is not clear to us why he thinks that this follows; as we have seen, it does not.) %%% Pitts seems to make this claim simply on the basis of the form of the Lagrangian. Is his claim true for the *Lagrangian-equivalent* Hamiltonian? That would make his argument question-begging rather than technically incorrect.

In fact, the same is true of all three cases (background charge distribution, charged particle, and minimal-coupled complex field): in the context of the extended Hamiltonian formalism, the charge distribution acts, and is acted on by, the objects defined by $(\vctr{A},\mu)$. $V$ is dynamically entirely decoupled from matter, and indeed in each case the dynamics for $V$ and its conjugate momentum are entirely trivial.

To see more directly how this replacement of $V$ by $\mu$ takes place, let's return to the background-field version of the extended Hamiltonian and ask: for given $\mu$ (and remembering that $\mu=\mu'+V$), what Lagrangian has the extended Hamiltonian as its Legendre transform? It is easy to confirm that the extended Hamiltonian is the Legendre transform of
\be 
L_{\mu'}[\vctr{A},V;\dot{\vctr{A}},\dot{V}]=\int \left\{\frac{1}{2}(\dot{\vctr{A}}-\nabla (V+\mu'))^2 -\frac{1}{2}(\nabla \times \vctr{A})^2 -( (V+\mu') \rho + \vctr{A} \cdot \vctr{J})\right\} 
\ee
It is immediately clear that the physics generated by $L_{\mu'}$ does not differ in any physically interesting way (and in particular, in any empirically accessible way) from that generated by the original Lagrangian (\ref{lag}). They differ by a mere labelling: what in one case is called $V$, in another case is called $V+\mu'$.

In this context, consider once more Pitts:
\begin{quote}
Velocities (such as appear in the electric field) are not physically recondite --- automobiles have gauges that measure them --- but canonical momenta are. 
\end{quote}
But of course the velocity \emph{of the vector potential} \vctr{A}, being gauge-dependent, is not measureable at all. \vctr{A}, and its velocity, is measureable only up to gauge transformations, which is to say that $\dot{\vctr{B}}$ is the only directly measureable velocity --- and 
 \vctr{B} and its time derivatives are invariant under the action of the constraints. The electric field is also measureable, via its dynamical interactions with charge, but in advance of checking the dynamics, we cannot assume that the thing being measured is a combination of $V$ and $\dot{\vctr{A}}$ rather than the momentum $\vctr{\pi}$. As we have seen, it is not, except in the special case where we set the Lagrange multiplier $\mu$ in the extended Hamiltonian equal to $V$.

\section{Conclusion}

``Constraints generate gauge transformations'' can mean, in electromagnetism, one of two things:
\begin{enumerate}
\item Both primary and secondary constraints generate transformations between gauge-related instantaneous states.
\item Arbitrary time-dependent combinations of the primary and secondary constraints generate transformations between gauge-related histories. %cannot 
\end{enumerate}
It is this second meaning that Pitts has in mind when he claims that the constraints generate bad physical transformations rather than gauge transformations. While arbitrary time-dependent combinations of the primary and secondary constraints do not generate transformations that are symmetries of the Hamiltonian formalism that is strictly equivalent to the original Lagrangian, they are symmetries of the extended Hamiltonian formalism. And the conventional wisdom about the extended Hamiltonian formalism---that it is empirically equivalent to the standard formalism---is correct: \emph{contra} Pitts' claim, the dynamical role of the electric field is played, within that formalism, by the momentum conjugate to the vector potential, and not by the combination $\dot{\vctr{A}}-\nabla V$ found in the unmodified Lagrangian formalism. Furthermore, the extended Hamiltonian formalism, for any given choice of secondary-constraint Lagrange multiplier, can be seen as the Legendre transform of a Lagrangian that is, transparently, empirically equivalent to the original electromagnetic Lagrangian.

%I can give you the .bib file if preferred

%%OP - that would make it easier for me to add references.

%\bibliographystyle{unsrt}
%\bibliography{../bib/general2}

\end{document}